\documentclass[preprint]{aastex}
\shorttitle{Proper Motion of the Large Magellanic Cloud}
\shortauthors{Pedreros et al.}
     
\begin{document}

\title{The Proper Motion of the Large Magellanic Cloud: A Reanalysis}

\author{Mario H. Pedreros\altaffilmark{1}} \affil{Departamento de F\'{\i}sica,
Facultad de Ciencias, Universidad de Tarapac\'{a}, Casilla 7-D, Arica, Chile}
\email{mpedrero@uta.cl,}

\and

\author{Edgardo Costa\altaffilmark{1} and Ren\'e
A. M\'endez\altaffilmark{1}} \affil{Departamento de Astronom\'{\i}a,
Universidad de Chile, Casilla 36-D, Santiago, Chile}
\email{[costa, rmendez]@das.uchile.cl}
     
\altaffiltext{1}{Visiting Astronomer, Cerro Tololo Inter-American
Observatory, National Optical Astronomy Observatories, operated by the
Association of Universities for Research in Astronomy, Inc. (AURA),\
under cooperative agreement with the National Science Foundation.}

\begin{abstract}

We have determined the proper motion (PM) of the Large Magellanic
Cloud (LMC) relative to four background quasi-stellar objects,
combining data from two previous studies made by our group, and new
observations carried out in four epochs not included the original
investigations.  The new observations provided a significant increase
in the time base and in the number of frames, relative to what was
available in our previous studies.  We have derived a total LMC PM
of $\mu$ = ($+$2.0$\pm$0.1) mas yr$^{-1}$, with a position angle of
$\theta$ = (62.4$\pm$3.1)$^\circ$.  Our new values agree well with most
results obtained  by other authors, and we believe we have clarified
the large discrepancy between previous results from our group.
Using published values of the radial velocity for the center of the
LMC, in combination with the transverse velocity vector derived from
our measured PM, we have calculated the absolute space velocity of
the LMC.  This value, along with some assumptions regarding the mass
distribution of the Galaxy, has in turn been used to calculate the
mass of the Milky Way.  Our measured PM also indicates that the LMC
is not a member of a proposed stream of galaxies with similar orbits
around our galaxy.

\end{abstract}

\section {INTRODUCTION}  

The present study is a follow-up of the works by Anguita, Loyola \&
Pedreros (2000, hereafter ALP) and Pedreros, Anguita \& Maza (2002,
hereafter PAM) in which the PM of the LMC was determined using the
"quasar method".  This method, fully described in ALP and PAM,
consists in using quasi-stellar objects (QSOs) in the background field
of the LMC, as fiducial reference points to determine its PM  In
this method, the position of the background QSOs is measured at
different epochs with respect to bona-fide field stars of the LMC
which define a local reference system (hereafter LRS). Because a QSO
can be considered a fiducial reference point, any motion detected will
will be a reflexion of the motion of the LRS of LMC stars.

As shown in Table 1, there is a rather large discrepancy, particularly
in Decl., between the PM of the LMC derived by ALP and that derived
by PAM, with ALP$-$PAM differences of $-$0.3 mas yr$^{-1}$
(1.5 $\sigma$) in R.A., and 2.5 mas yr$^{-1}$ (12.5 $\sigma$) in Decl.
This difference prompted us to add new epochs to our database (using
the same equipment and set-up used by ALP and PAM) and to make a full
reanalysis of the entire data set.

Here we report the results obtained combining data from previous
studies by our group, with new observations carried out in three
additional epochs (not included in the original investigation), for
the LMC quasar fields Q0459-6427, Q0557-6713, Q0558-6707, and
Q0615-6615 (in the same nomenclature used by ALP and PAM). The
original study of field Q0459-6427 was reported in PAM, and those of
Q0557-6713, Q0558-6707 and Q0615-6615 in ALP.  As can be seen in Table
2, which summarizes the total observational material used in the
present paper, our new data provides a significant increase, in
time base and in the number of frames, relative to what was available
in ALP and PAM.  The increase in time base for the fields Q0459-6427,
Q0557-6713, Q0558-6707 and Q0615-6615 was 19\%, 65\%, 126\% and 65\%,
respectively.  The corresponding increase in data points was 7\%,
18\%, 59\% and 56\%, respectively.

\section {OBSERVATIONS AND REDUCTIONS}

The new observations were carried out with a 24$\mu$ pixels Tektronix
1024x1024 CCD detector attached to the Cassegrain focus of the CTIO
1.5 m telescope in its f/13.5 configuration (scale: 0.24 $\arcsec$/pixel).
Only astrometric observations were
secured.  Because for each QSO field we adopted the same LRS used by
ALP or PAM, there was no need for additional photometric observations.
Finding charts for the reference stars and the
background QSO in each field can be found in ALP or PAM.  As was done
in our previous studies, the astrometric observations were made using a
Kron-Cousins {\it R}-band filter, in order to minimize differential
color refraction effects.

The method used for the determination of the LMC's PM is the same as that
explained in ALP and PAM.  Only data not included in those two
previous studies went through the full reduction procedure.  For data
already included in those studies, we used the available raw
coordinates for the centroids of the reference stars and background
QSOs. Both, the existing and the newly determined raw coordinates, were
treated by means of the same custom programs used in PAM.

In brief, the (x,y) coordinates of the QSO and the LMC field reference
stars in each image were determined using the DAOPHOT package (Stetson
1987), and then corrected for differential color refraction and
transformed to barycentric coordinates.  Then, by averaging the
barycentric coordinates of the best set of consecutive images taken of
each QSO field throughout our program, a standard reference frame
(SRF) was defined for every field.  All images, taken at different
epochs, of each field, were then referred to its corresponding SRF.
This was done through multiple regression analysis by fitting both
sets of coordinates to quadratic equations of the form: $X = a_0 +
a_1x + a_2y + a_3x^2;~~~~Y = b_0 + b_1x + b_2y + b_3x^2$; where
($X,Y$) are the coordinates on the SRF system and ($x,y$) are the the
observed barycentric coordinates.  It was found that the above
transformation equations yielded the best results for the registration
into the SRF, showing no remaining systematic trends in the data.

\section {RESULTS}

Tables 3-6 list the residual PM (relative to the barycenter of the
field's SRF) and photometry (this latter from ALP or PAM, and included
here for completeness) of the stars defining the LRS in each of our
four QSO fields. Star IDs are the same as those in PAM and ALP, for the
corresponding fields.  The PM uncertainties correspond to the error in
the determination of the slope of the best-fit line. Inspection of
these tables shows that the PM uncertainty of most of the reference
stars is comparable to, or larger than, their derived PM value,
implying that these PM do not represent internal motions  in the LMC.

In Figure 1 we present the PM maps for the reference stars listed in
Tables 3-6.  The dispersion around the mean turned out to be $\pm$0.34,
$\pm$0.79, $\pm$0.54, and $\pm$0.41 mas yr$^{-1}$ in R.A., and $\pm$0.52,
$\pm$0.71, $\pm$0.58, $\pm$0.62 mas yr$^{-1}$ in
Decl., for Q0459-6427, Q0557-6713, Q0558-6707 and Q0615-6615,
respectively.  Based on the above argument, the scatter seen in the
plots probably stems entirely from the random errors in the
measurements, and does not represent the actual velocity dispersion in
the LMC.

In Figure 2 we present position $vs.$ epoch diagrams for the QSO
fields in R.A.  ($\Delta\alpha$cos$\delta$) and
Decl. ($\Delta\delta$), were $\Delta\alpha$cos$\delta$ and
$\Delta\delta$ represent the positions of the QSOs on different CCD
frames, relative to the barycenter of the SRF. These diagrams were
constructed using individual position data for the QSO in each CCD
image as a function of epoch. In Table 7 we give, for each epoch,
the mean barycentric positions of the QSOs along with their mean
errors, the number of points used to calculate the mean for each
coordinate, and the CCD detectors used.  Symbol sizes in Figure 2 are
proportional to the number of times the measurements yielded the same
coordinate value for a particular epoch. 
The best-fit straight lines resulting from
simple linear regression analysis on the data points are also shown.
The negative values of the line slopes correspond to the measured PM
of the barycenter of the LRS, in each QSO field, relative to the SRF.

Table 8 summarizes our results for the measured PM of the LMC.  Column (1)
gives the quasar identification, columns (2) and (3) the R.A. and Decl.
components of the LMC's PM (together with their standard deviations)
respectively, and, finally, columns (4), (5) and (6) the
number of frames, the number of epochs, and the observation period,
respectively.  It should be noted that the rather small quoted errors
for the PM come out directly from what the least-square fit yields
as the uncertainty in the determination of the slope of the best fit
line.

\section {COMPARISON TO OTHER PROPER MOTION WORK}

Table 9 lists the results of all available measurements of the LMC's PM
having uncertainties smaller than 1 mas yr$^{-1}$ in both components,
as well as the reference system used in each case. With the exception
of those cases noted as "Field" in the first column, all the PM listed
in Table 9 are relative to the LMC's center. To facilitate comparisons, we present
our current results in both ways.  As explained in the next section, our
PM values relative to the LMC's center were obtained correcting the
field PM for the rotation of the plane of the LMC.

Our results are in reasonable agreement with most of the available data.
They agree particularly well with those of Kroupa et al. (1994), who used
the Positions and Proper Motions Star Catalog (PPM, R\"{o}ser et al., 1993)
as reference system, and also with the HST unpublished result of Kallivayalil
et al. (2005), who used QSOs as reference system.  On the other hand, there still
is a significant discrepancy with ALP's result in Decl.  We will further discuss
this issue in \S 6.

In Table 9 we have not included a recent determination of the LMC's
PM by Momany \& Zaggia (2005) using the USNO CCD Astrograph all-sky
Catalog (UCAC2, Zacharias et al. 2004), because, as confirmed by the
errors declared by the authors themselves ($\sim$3~mas in both
coordinates), the internal accuracy of their methodology is not
comparable with ours.  Numerous tests carried out by our group, favor
the use of fiducial reference points in combination with a LRS defined
by relatively few, well studied (bona-fide members, free of contamination
from neighboring stars, good signal-to-noise, etc.) LMC stars, to 
determine a PM of this nature.  Interestingly, their result
[$\mu_{\alpha}$cos$\delta$,$\mu_{\delta}$] $\sim$ [+0.84,+4.32] mas yr$^{-1}$
is in reasonable agreement with that of ALP.

Combining the components given in the last entry of Table 9, we derive
a total LMC PM of $\mu$ = ($+$2.0$\pm$0.1) mas yr$^{-1}$, with a position
angle of $\theta$ = (62.4$\pm$3.1)$^\circ$, measured eastward from the meridian
joining the center of the LMC to the north celestial pole.  This result is
compatible (particularly the PM's absolute value) with theoretical models
(Gardiner et al. 1994), which predict
a PM for the LMC in the range 1.5$-$2.0 mas yr$^{-1}$, with a position
angle of \mbox{$\theta\approx 90^\circ$}.

\section {SPATIAL VELOCITY OF THE LMC AND MASS OF THE GALAXY}

Using the PM of the LMC determined in \S 3, and the radial velocity
of the center of the LMC (adopted from the literature), we can
calculate the radial and transverse components of the velocity for the
LMC, as seen from the center of the Galaxy, along with other parameters 
described below.  To do this we basically followed the procedure
outlined by Jones et al. (1994).  In the calculations we used as basic
LMC parameters those given in Table 8 of ALP, and assumed a rotational
velocity v$_{\Phi}$ = 50 km s$^{-1}$ and a radial velocity
V$_r$ = 250 km s$^{-1}$ for the LMC.\\

In order to determine, from our measured PM values, the space velocity
components of the LMC, and its PM with respect to the Galactic Rest Frame
(GRF), a series of steps were required.  These include:
1. A correction to our measured PM values to account for the rotation of
the plane of the LMC;  2. A transformation of the corrected PM into transverse
velocity components with respect the the center of the LMC, the Sun, the LSR
and the center of the Galaxy; both in the equatorial and galactic coordinate systems.
These transverse velocities, in combination with the radial velocity of the center
of the LMC (adopted from the literature), allowed us to derive the components of the 
space velocity of the LMC corrected for the Sun's peculiar motion relative to the LSR,
and also corrected for the velocity of the LSR itself, relative to the center of the Galaxy.
The above calculations were made using an $ad-hoc$ computer program, developed by one of the
authors (MHP), which generates results consistent with those from an independent software
(Piatek, 2005; private communication).\\

The results of the above procedure applied to our four quasar fields are presented in
Table 10.  In rows 1-2 we list the  R.A. and Decl. corrections to our measured PM
to account for the rotation of the plane of the LMC, and in rows 3-4, the corresponding
corrected PM values, in equatorial coordinates, as viewed by an observer located at
the center of the LMC.
In rows 5-8 we
give calculated PM values relative to the GRF, both in equatorial and galactic coordinates.
These values correspond to the LMC's PM as seen by an observer located at the Sun, with
the contributions to the PM, from the peculiar solar motion and from the LSR's motion, removed.
In rows 9-11 we give the $\Pi$, $\Theta$ and $Z$ components of the space velocity in a
rectangular cartesian coordinate system centered on the LMC (as defined by Schweitzer et al.,
1995, for the Sculptor dSph).  The $\Pi$ component is parallel to the projection onto the
Galactic plane of the radius vector from the center of the Galaxy to the center of the LMC,
and is positive when it points radially away form the Galactic center.
The $\Theta$ component is perpendicular to the $\Pi$ component, parallel to 
the Galactic plane, and points in the direction of rotation of the Galactic disk.
The $Z$ component points in the direction of the Galactic north pole. These three components
are free from the Sun's peculiar motion and LSR motion. In rows 12-13 we give the LMC's radial
and transverse space velocities, as seen by an hypothetical observer located at the center of
the Galaxy, and at rest with respect to the Galactic center.\\

All of the above calculations were carried out assuming a distance of 50.1 kpc of the LMC
from the Sun, a distance of 8.5 kpc of the Sun from the Galactic center, a 220 km s$^{-1}$ 
circular velocity of the LSR and a peculiar velocity of the Sun relative to the LSR of
(u$_{\sun}$,v$_{\sun}$,w$_{\sun}$) = ($-$10,5.25,7.17) km s$^{-1}$ (Dehnen \& Binney 1998),
These components are positive if u$_{\sun}$ points radially away from the Galactic center,
v$_{\sun}$ points in the direction of Galactic rotation and w$_{\sun}$ is directed towards
the Galactic north pole.\\

Although the matter was not addressed here, the values presented in table 10 can be used to
determine the orbit of the LMC and therefore study possible past and future interactions
of the LMC with other Local Group galaxies.\\

If we assume that the LMC is gravitationally bound to, and in an elliptical orbit, around the
Galaxy, and that the mass of the Galaxy is contained within 50 kpc of the galactic center, we    
can make an estimate of the lower limit of its mass through the expression:     
     
\begin{displaymath}     
{\rm M_{G}} = ({\rm r_{LMC}} / 2 {\rm G})[{\rm V^2_{gc,~r}} +      
{\rm V^2_{gc,~t}}~(1 - {\rm r}_{\rm LMC}^2 / {\rm r^2_a})]     
/ (1-{\rm r_{LMC}} / {\rm r_a})       
\end{displaymath}     
     
\noindent     
where r$_{\rm a}$ is the LMC's apogalacticon distance and r$_{\rm LMC}$ its present distance.

For r$_{\rm a}$ = 300 kpc (Lin et al. 1995) we obtain $\rm M_{G}$ values of :     
(8.2 $\pm$ 1.3), (9.9$\pm$ 1.6), (3.0 $\pm$ 0.8)~and~(12 $\pm$ 2) $\times 10^{11} \cal M_{\sun}$,
for the fields, Q0459-6427, Q0557-6713, Q0558-6707 and Q0615-6615, 
respectively. The above values result in a weighted average of: $\langle
M_{G}\rangle = (5.9 \pm 0.6) \times 10^{11} \cal M_{\sun}$ for the
estimated mass of our Galaxy enclosed within 50 kpc.\\
       
To evaluate the effect of the rotational velocity of the LMC on the determination
of the mass of our galaxy, we also carried out calculations using the extreme
values v$_{\Phi}$ = 0 km s$^{-1}$ (zero rotation) and v$_{\Phi}$ = 90 km s$^{-1}$.
The weighted mass
averages for 0 and 90 km s$^{-1}$ resulted to be $(5.6 \pm 0.6)\times 10^{11}$
and $(6.3\pm 0.6)\times 10^{11} \cal M_{\sun}$, respectively.  Our results 
are summarized in Table 11.\\

It should be noted that (although slightly larger), all our values for $\rm M_{G}$
are compatible with the recent theoretical $5.5 \times 10^{11} \cal M_{\sun}$
upper mass limit of the Galaxy given by Sakamoto et al. (2003). They
are also compatible with the assumption that the LMC is bound to to the Galaxy.

\section{DISCUSSION}

\subsection{The ALP-PAM Discrepancy}

Given the implications of the result obtained by ALP for the PM of the
LMC, in relation to our understanding of the interactions between the
Galaxy and the Magellanic Clouds (see, e.g, Momany \& Zaggia, 2005), and
the reality of streams of galaxies with similar orbits
around the Galaxy (see, e.g, Piatek et al., 2005), the main objective
of the present work was to clarify the discrepancy between the
previous determinations of the PM of the LMC by our group: the
"ALP-PAM Discrepancy".  In this section, we further elaborate on some
of the thoughts originally proposed in PAM, in order to explain the
discrepancy of ALP, originally with PAM, and now also with the new
result presented in this paper.

First, the fact that the observations used here were made
with essentially the same equipment and instrumental set-up
as those by ALP, precludes any arguments relating the observed
discrepancy to the existence of systematic errors in the observational
data.  Such errors would affect our data in the same way way as those
of ALP. 

Second (as explained in \S 2), in the reduction process of the ALP
and PAM data incorporated in the present work we adopted the same QSO
and reference stars centroid coordinates (x,y) used in those works.
Furthermore, the new data included in the present calculations was
processed using the same procedure used in ALP to obtain the (x,y)
coordinates. Therefore, the centroid coordinates should not be a
source of a systematic error either.

The subsequent procedures to obtain the PM were also basically the
same, the sole exception being the inclusion of a quadratic term in
the transformation equations used for the registration (also included
in PAM's equations, but not in ALP's).  Tests carried out using ALP's
data alone showed however that the effect of including quadratic terms
is marginal (as was suspected), and does not account for the observed
discrepancy.

Considering that our current result -which includes re-processed data
from ALP- agrees quite well with measurements by other groups, we
conclude that ALP's results might be affected by an unidentified systematic
error in Decl.  Since in the present work we used ALP's unmodified
(x,y) coordinates, we believe that this error could have originated
in the processing of the Decl. PM instead of the coordinates
themselves. 

It should be pointed out that the UCAC2-based result from
Momany \& Zaggia (2005), which is consistent with that from ALP, is
currently also considered to be affected by an as yet unidentified
systematic error (Momany \& Zaggia, 2005; Kallivayalil et al., 2005).

We would finally like to note that our new result for field Q0459-6427
is consistent with PAM.

\subsection{Membership of the LMC to a Stream}

Lynden-Bell \& Lynden-Bell (1995), have proposed that the LMC, together
with the SMC, Draco and Ursa Minor, and possibly Carina and Sculptor,
define a stream of galaxies with similar orbits around our galaxy.
Their models predict a PM for each of member of the stream, which can
be compared to their measured PM to evaluate the reality of the stream.

For the LMC they predict PM components of 
[$\mu_{\alpha}$cos$\delta$,$\mu_{\delta}$] = [+1.5,0] mas yr$^{-1}$,
giving a total PM of $\mu$ = +1.5 mas yr$^{-1}$, with a position angle
of $\theta$ = 90$^\circ$.  A comparison of this prediction with our result
[$\mu$ = ($+$2.0$\pm$0.1) mas yr$^{-1}$, $\theta$ = (62.4$\pm$3.1)$^\circ$],
shows that our measured values of $\mu$ and $\theta$ are, respectively,
5.1$\sigma$ and 8.9$\sigma$ away from the predicted values.  This result
indicates that the LMC does not seem to be a member of the above stream
(it is worth mentioning that Piatek et al. (2005), using HST data, have
concluded that Ursa Minor is not a member of this stream).\\

\newpage
\acknowledgments

MHP is greatful of the support by the Universidad de Tarapac\'{a}
research fund (project \# 4722-02).  EC and RAM acknowledge support by
the Fondo Nacional de Investigaci\'on Cient\'{\i}fica y Tecnol\'ogica
(proyecto No. 1050718, Fondecyt) and by the Chilean Centro de
Astrof\'{\i}sica FONDAP (No. 15010003). It is also a pleasure to thank
T. Mart\'\i nez for helping with data processing.

We would like to thank the referee, Dr. S. Piatek, for his constructive
comments.

\clearpage

{\bf Figure Captions}

{\bf Figure 1.}  Residual proper motion maps for the reference stars
listed in Tables 3-6.  The dispersion around the mean is $\pm$ 0.34,
$\pm$ 0.79, $\pm$ 0.54, and $\pm$ 0.41 mas yr$^{-1}$ in R.A., and
$\pm$ 0.52, $\pm$ 0.71, $\pm$ 0.58, $\pm$ 0.62 mas yr$^{-1}$ in Decl.,
for Q0459-6427, Q0557-6713, Q0558-6707 and Q0615-6615, respectively.\\

{\bf Figure 2a.}  Relative positions in Right Ascension
($\Delta\alpha$cos$\delta$) $vs.$ epoch of observation for the studied
fields. The values of $\Delta\alpha$cos$\delta$ represent the
individual positions of the QSO on different CCD frames relative to
the barycenter of the SRF. Symbol sizes are proportional to the number
of times the measurements yielded the same coordinate value for a
particular epoch (extra small, small, medium, large, and extra large
sizes indicate 1 through 5 measurements per epoch, respectively).  The
best-fit straight lines from linear regression analyses on the data
are also shown.\\

{\bf Figure 2b.}  Relative positions in declination ($\Delta\delta$)
$vs.$ epoch of observation for the studied fields. The values of
$\Delta\delta$ represent the individual positions of the QSO on
different CCD frames relative to the barycenter of the SRF. Symbol
sizes and best-fit straight lines as described in Fig 2a.\\

\clearpage
\begin{figure}
\epsscale{0.99}
\plotone{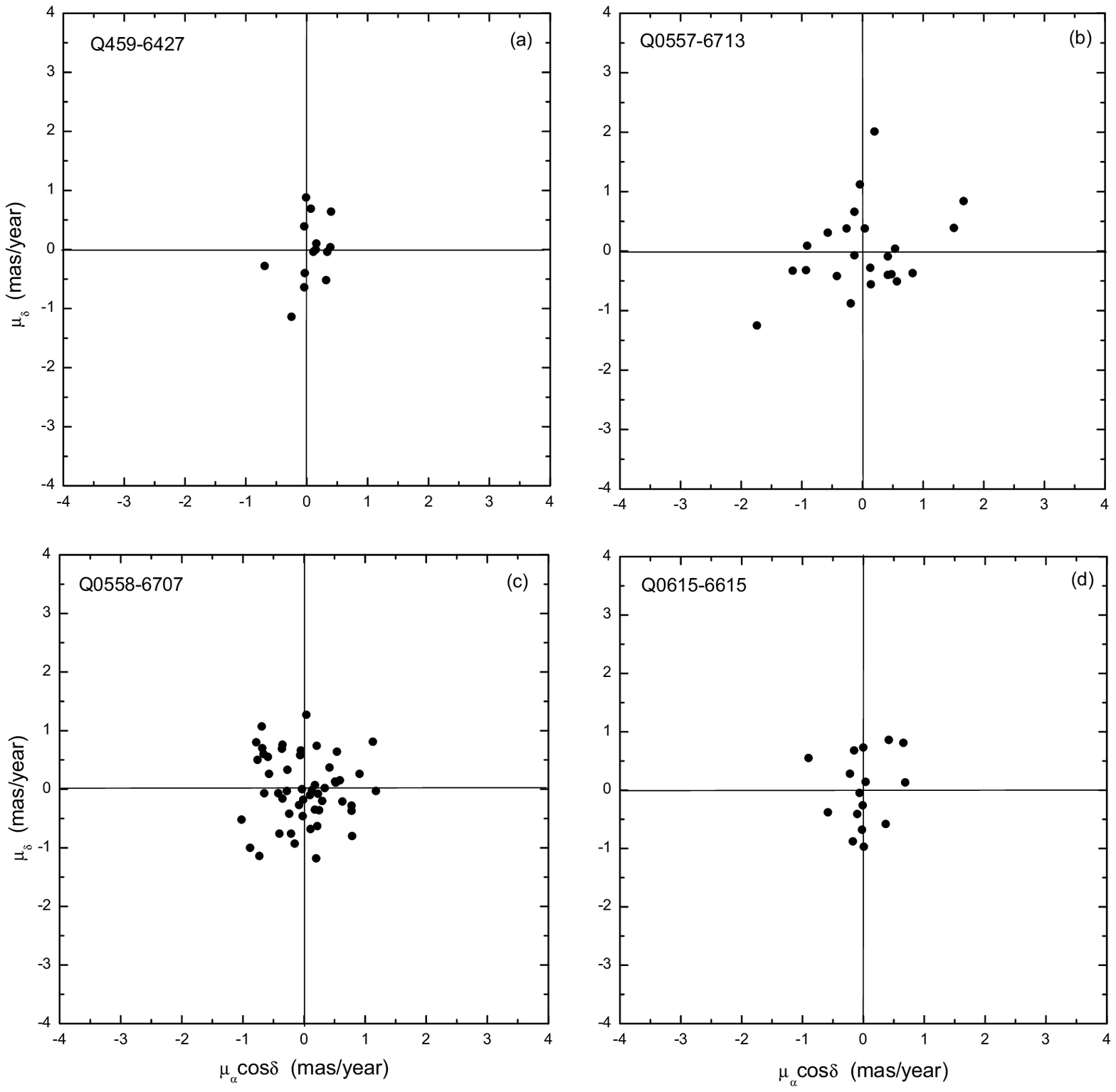}
\caption{}
\end{figure}

\clearpage
\begin{figure}
\epsscale{0.99}
\plotone{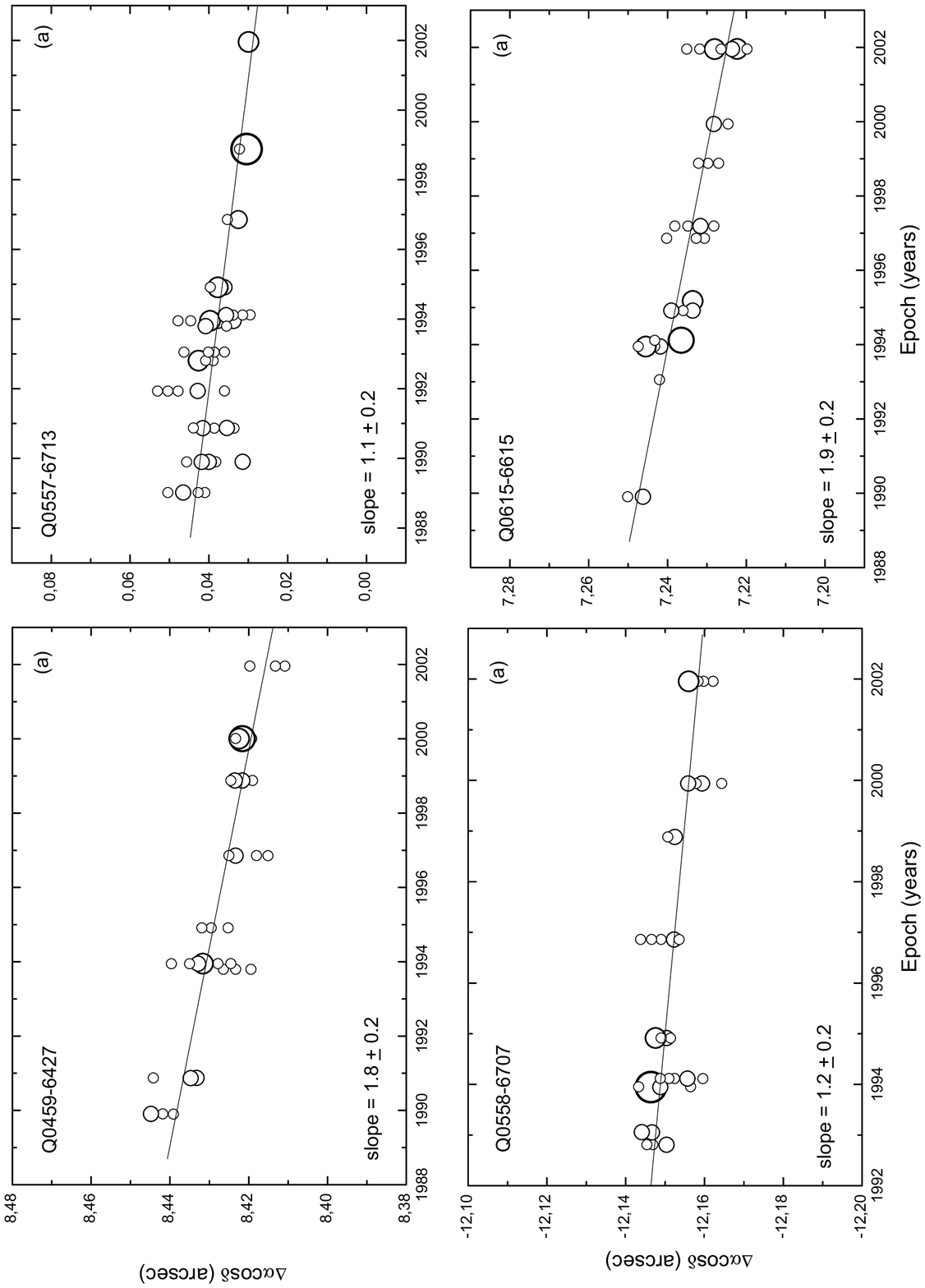}
\caption{}
\end{figure}

\clearpage
\begin{figure}
\epsscale{0.99}
\plotone{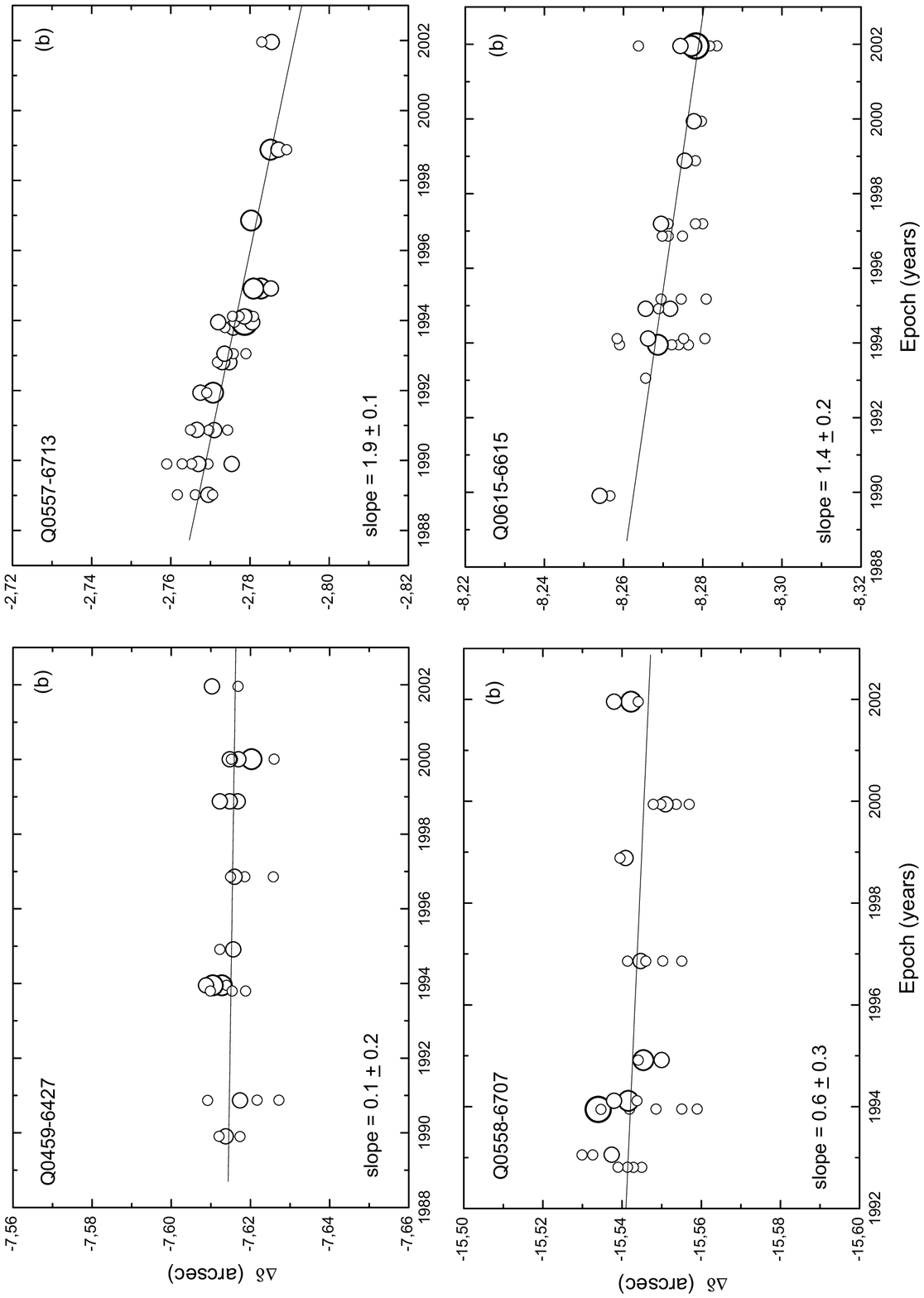}
\caption{}
\end{figure}

\clearpage
     
\begin{table*}     
\begin{flushleft}     
\begin{tabular}{lcccc}     
\multicolumn{4}{c}{Table 1} \\
     \multicolumn{4}{c}{Previous determinations of the LMC proper motion
using the quasar method}\\

\noalign{\smallskip}     
\hline \hline     
\noalign{\smallskip}     
{Source}& $\mu_{\alpha}$cos$\delta$ & {$\mu_{\delta}$}&{Weighted Mean from}\\     
&  mas yr$^{-1}$  & mas yr$^{-1}$  & & \\     
\noalign{\smallskip}     
\hline     
\noalign{\smallskip}

ALP(LMC center) &$+$1.7 $\pm$ 0.2 &$+$2.9 $\pm$ 0.2 &3 fields\\
PAM (LMC center) &$+$2.0 $\pm$ 0.2 &$+$0.4 $\pm$ 0.2 &1 field\\

\hline      
\end{tabular}     
\end{flushleft}     
\end{table*}       
  
\begin{table*}
\begin{flushleft}
\begin{tabular}{cccccccc}  
\multicolumn{8}{c}{Table 2} \\ 
\multicolumn{8}{c} {Observational material for the LMC QSO fields}\\
\hline \hline
Field & & & Old Data  & & & New Data   \\
& Source & Epochs & \# Frames & Epoch Range & Epochs & \# Frames
& Epoch Range \\ 
\hline
Q0459-6427& PAM& 8& 44 & 1989.91$-$2000.01 & 1 & 3 & 2001.96\\
Q0557-6713& ALP& 11& 61 & 1989.02$-$1996.86 & 2 & 11 &
1998.88$-$2001.96\\
Q0558-6707& ALP& 6& 32 & 1992.81$-$1996.86 & 3 & 19 &
1998.88$-$2001.96\\
Q0615-6615& ALP& 8& 32 & 1989.90$-$1997.19 & 3 & 18 &
1998.88$-$2001.96\\

\hline
\end{tabular}
\end{flushleft}  
\end{table*}

\begin{table*}
\begin{flushleft}
\begin{tabular}{cccccccc}  
\multicolumn{8}{c}{Table 3} \\ 
\multicolumn{8}{c} {Local Reference System for the Q0459$-$6427 field}\\
\hline \hline
Star$^{(a)}$& $\mu_{\alpha}$cos$\delta$ & $\sigma$ & $\mu_{\delta}$ & $\sigma$ & V &B$-$V
 &V$-$R \\
ID &mas yr$^{-1}$&mas yr$^{-1}$ & mas yr$^{-1}$ & mas yr$^{-1}$ &mag & mag  
& mag \\ 
\hline    

1&      $ $~~0.0 & 0.3&   $+$0.9 & 0.2&   18.71&  0.95&   0.52\\
2&      $-$0.7 & 0.3&   $-$0.3 & 0.4&   19.01&  0.67&   0.38\\
3&      $+$0.6 & 0.2&   $-$0.4 & 0.2&   19.02&  0.86&   0.47\\
4&      $ $~~0.0 & 0.3&   $ $~~0.0 & 0.3&   18.88&  0.96&   0.52\\
5&      $ $~~0.0 & 0.3&   $+$0.6 & 0.3&   18.71&  0.98&   0.54\\
6&      $-$0.1 & 0.2&   $-$0.4 & 0.2&   18.22&  1.03&   0.58\\
7&      $+$0.1 & 0.2&   $+$0.1 & 0.2&   18.08&  1.03&   0.57\\
8&      $ $~~0.0 & 0.5&   $-$0.1 & 0.4&   17.98&  0.89&   0.52\\
9&      $+$0.1 & 0.3&   $-$0.6 & 0.4&   19.18&  0.84&   0.43\\
10&     $+$0.3 & 0.1&   $ $~~0.0 & 0.2&   17.94&  1.15&   0.63\\
11&     $ $~~0.0 & 0.3&   $+$0.4 & 0.3&   18.64&  0.91&   0.50\\
12&     $-$0.3 & 0.3&   $-$1.2 & 0.3&   19.03&  0.88&   0.48\\
13&     $+$0.4 & 0.3&   $+$0.6 & 0.3&   18.98&  0.86&   0.48\\
14&     $-$0.7 & 0.3&   $+$0.1 & 0.3&   18.66&  0.23&   0.03\\
15&     $+$0.2 & 0.1&   $-$0.1 & 0.2&   17.70&  1.08&   0.59\\
16&     $+$0.4 & 0.2&   $+$0.3 & 0.2&   16.70&  1.43&   0.82\\
17&     $-$0.3 & 0.2&   $+$0.3 & 0.2&   19.17&  0.95&   0.51\\

\hline
\end{tabular}

\small{$ ^{(a)}$Star IDs are the same as those in PAM}\\
\end{flushleft}  
\end{table*}         

\begin{table*}
\begin{flushleft}
\begin{tabular}{cccccccc}  
\multicolumn{8}{c}{Table 4} \\ 
\multicolumn{8}{c} {Local Reference System for the Q0557$-$6713 field}\\
\hline \hline
Star$^{(a)}$&  $\mu_{\alpha}$cos$\delta$ & $\sigma$ & $\mu_{\delta}$ & $\sigma$ & V &B$-$V
 &V$-$R \\
ID &mas yr$^{-1}$&mas yr$^{-1}$ & mas yr$^{-1}$ & mas yr$^{-1}$ &mag & mag  
& mag \\  
\hline    

1&      $-$0.1 & 0.1&   $-$0.1 & 0.2&   17.07&  $-$0.07&        $-$0.04\\
2&      $+$0.5 & 0.2&   $ $~~0.0 & 0.1&   17.75&  1.14&           0.56\\
3&      $-$0.1 & 0.2&   $+$0.7 & 0.2&   18.35&  0.84&           0.45\\
4&      $+$0.2 & 0.3&   $+$2.0 & 0.3&   18.64&  0.68&           0.38\\
5&      $-$0.2 & 0.4&   $-$0.9 & 0.2&   16.93&  1.13&           0.55\\
6&      $+$0.6 & 0.2&   $-$0.5 & 0.2&   17.72&  1.22&           0.62\\
7&      $-$0.9 & 0.3&   $-$0.3 & 0.3&   18.73&  1.09&           0.48\\
8&      $-$1.2 & 0.2&   $-$0.3 & 0.2&   17.29&  0.83&           0.46\\
9&      $+$1.5 & 0.3&   $+$0.4 & 0.6&   18.52&  1.00&           0.52\\
10&     $-$1.7 & 0.6&   $-$1.2 & 0.3&   18.28&  0.00&           $-$0.05\\
11&     $+$0.4 & 0.2&   $-$0.4 & 0.2&   17.34&  1.17&           0.56\\
12&     $-$0.6 & 0.3&   $+$0.3 & 0.2&   18.66&  1.00&           0.48\\
13&     $-$0.4 & 0.2&   $-$0.4 & 0.2&   18.23&  0.75&           0.37\\
14&     $+$1.7 & 0.3&   $+$0.8 & 0.3&   18.13&  0.82&           0.42\\
15&     $+$0.8 & 0.3&   $-$0.4 & 0.2&   18.48&  0.80&           0.43\\
16&     $+$0.4 & 0.2&   $-$0.1 & 0.2&   18.26&  1.09&           0.53\\
17&     $+$0.1 & 0.2&   $-$0.6 & 0.2&   17.78&  0.95&           0.51\\
18&     $+$0.5 & 0.4&   $-$0.4 & 0.3&   17.57&  $-$0.12&        $-$0.06\\
19&     $+$0.1 & 0.1&   $-$0.3 & 0.2&   17.21&  1.19&           0.63\\
20&     $-$0.9 & 0.3&   $+$0.1 & 0.3&   18.69&  0.99&           0.50\\
21&     $-$0.3 & 0.1&   $+$0.4 & 0.2&   17.30&  0.77&           0.38\\
22&     $ $~~0.0 & 0.3&   $+$1.2 & 0.3&   18.05&  $-$0.08&        $-$0.08\\
23&     $ $~~0.0 & 0.1&   $+$0.4 & 0.2&   16.23&  $-$0.17&        $-$0.09\\

\hline
\end{tabular}

\small{$ ^{(a)}$Star IDs are the same as those in ALP}\\
\end{flushleft}  
\end{table*}         

\begin{table*}
\begin{flushleft}
\begin{tabular}{cccccccc}  
\multicolumn{8}{c}{Table 5} \\ 
\multicolumn{8}{c} {Local Reference System for the Q0558$-$6707 field}\\
\hline \hline
Star$^{(a)}$&  $\mu_{\alpha}$cos$\delta$ & $\sigma$ & $\mu_{\delta}$ & $\sigma$ & V &B$-$V
 &V$-$R \\
ID &mas yr$^{-1}$&mas yr$^{-1}$ & mas yr$^{-1}$ & mas yr$^{-1}$ &mag & mag  
& mag \\  
\hline    

1&      $+$0.8 & 0.5&   $-$0.8 & 0.7&   18.94&  0.84&   0.44\\
2&      $-$0.7 & 0.4&   $-$1.1 & 0.4&   16.44&  1.78&   0.91\\
3&      $+$0.2 & 0.3&   $-$1.2 & 0.4&   17.88&  0.90&   0.46\\
4&      $+$0.8 & 0.5&   $-$0.4 & 0.6&   18.94&  0.85&   0.46\\
5&      $-$0.9 & 0.3&   $-$1.0 & 0.7&   19.01&  0.90&   0.44\\
6&      $-$0.7 & 0.2&   $+$0.6 & 0.2&   18.30&  0.88&   0.49\\
7&      $+$0.5 & 0.2&   $+$0.1 & 0.3&   17.78&  1.18&   0.62\\
8&      $+$1.1 & 0.4&   $+$0.8 & 0.4&   18.36&  ....&$-$0.11\\
9&      $+$0.2 & 0.2&   $-$0.4 & 0.2&   17.39&  1.34&   0.70\\
10&     $+$0.5 & 0.2&   $+$0.1 & 0.3&   18.43&  0.86&   0.46\\
11&     $-$0.8 & 0.2&   $+$0.8 & 0.2&   17.79&  1.13&   0.59\\
12&     $ $~~0.0 & 0.2&   $ $~~0.0 & 0.3&   18.59&  0.88&   0.45\\
13&     $-$0.2 & 0.3&   $-$0.4 & 0.4&   18.34&  -0.02&0.00\\
14&     $ $~~0.0 & 0.4&   $+$0.7 & 0.4&   18.20&  0.01&$-$0.01\\
15&     $-$0.6 & 0.2&   $+$0.6 & 0.2&   17.44&  1.26&   0.66\\
16&     $-$0.7 & 0.4&   $+$0.7 & 0.5&   19.00&  0.91&   0.49\\
17&     $-$0.7 & 0.3&   $+$1.1 & 0.3&   18.48&  0.69&   0.40\\
18&     $-$0.1 & 0.5&   $+$0.6 & 0.6&   18.98&  0.90&   0.48\\
19&     $+$0.2 & 0.3&   $+$0.7 & 0.4&   18.32&  -0.13&$-$0.02\\
20&     $-$0.4 & 0.4&   $+$0.8 & 0.5&   19.00&  0.87&   0.49\\
21&     $+$0.4 & 0.3&   $+$0.4 & 0.5&   18.84&  0.91&   0.48\\
22&     $-$0.3 & 0.4&   $+$0.3 & 0.4&   18.83&  0.91&   0.48\\
23&     $-$0.4 & 0.4&   $-$0.2 & 0.2&   16.29&  0.02&   0.17\\
24&     $ $~~0.0 & 0.2&   $-$0.2 & 0.2&   17.56&  1.27&   0.67\\
25&     $+$0.2 & 0.2&   $-$0.1 & 0.3&   17.69&  1.15&   0.60\\
\hline
\end{tabular}

\small{$ ^{(a)}$Star IDs are the same as those in ALP}\\
\end{flushleft}  
\end{table*}         

\begin{table*}
\begin{flushleft}
\begin{tabular}{cccccccc}  
\multicolumn{8}{c}{Table 5} \\ 
\multicolumn{8}{c} {Local Reference System for the Q0558$-$6707 field(continued)}\\
\hline 
Star$^{(a)}$&  $\mu_{\alpha}$cos$\delta$ & $\sigma$ & $\mu_{\delta}$ & $\sigma$ & V &B$-$V
 &V$-$R \\
ID &mas yr$^{-1}$&mas yr$^{-1}$ & mas yr$^{-1}$ & mas yr$^{-1}$ &mag & mag  
& mag \\  
\hline 

26&     $+$0.2 & 0.2&   $+$0.1 & 0.4&   18.72&  1.20&   0.57\\
27&     $-$0.1 & 0.2&   $-$0.3 & 0.3&   18.66&  1.00&   0.54\\
28&     $-$0.6 & 0.2&   $+$0.3 & 0.2&   17.31&  1.25&   0.64\\
29&     $+$0.6 & 0.4&   $-$0.2 & 0.4&   18.92&  0.89&   0.47\\
30&     $-$0.4 & 0.3&   $-$0.1 & 0.3&   18.18&  1.25&   0.59\\
31&     $+$0.8 & 0.3&   $-$0.3 & 0.3&   18.55&  1.01&   0.54\\
32&     $+$0.2 & 0.3&   $-$0.4 & 0.3&   18.07&  1.12&   0.61\\
33&     $+$0.6 & 0.2&   $+$0.2 & 0.2&   17.12&  1.46&   0.76\\
34&     $+$0.9 & 0.8&   $+$0.3 & 0.8&   18.68&  0.84&   0.48\\
35&     $+$0.2 & 0.2&   $-$0.6 & 0.2&   17.42&  0.90&   0.47\\
36&     $+$0.1 & 0.3&   $-$0.7 & 0.4&   18.75&  0.83&   0.45\\
37&     $-$0.4 & 0.3&   $-$0.8 & 0.2&   18.65&  1.27&   0.56\\
38&     $ $~~0.0 & 0.2&   $-$0.5 & 0.2&   17.89&  1.18&   0.60\\
39&     $+$0.1 & 0.4&   $-$0.1 & 0.4&   19.12&  0.92&   0.50\\
40&     $-$0.7 & 0.4&   $-$0.1 & 0.4&   19.05&  0.88&   0.50\\
41&     $+$0.3 & 0.2&   $-$0.2 & 0.4&   18.35&  0.01&   0.02\\
42&     $-$0.2 & 0.2&   $-$0.9 & 0.5&   18.58&  0.89&   0.50\\
43&     $+$0.1 & 0.2&   $ $~~0.0 & 0.2&   17.45&  1.32&   0.68\\
44&     $+$0.3 & 0.5&   $ $~~0.0 & 0.6&   19.01&  0.85&   0.49\\
45&     $ $~~0.0 & 0.3&   $+$1.3 & 0.3&   18.46&  0.66&   0.43\\
46&     $-$0.8 & 0.3&   $+$0.5 & 0.4&   19.05&  0.86&   0.52\\
47&     $-$0.4 & 0.4&   $+$0.7 & 0.6&   19.04&  1.01&   0.54\\
48&     $+$0.5 & 0.2&   $+$0.6 & 0.3&   16.81&  0.08&   0.06\\
49&     $+$1.2 & 0.4&   $ $~~0.0 & 0.3&   19.04&  0.83&   0.49\\
50&     $-$1.0 & 0.3&   $-$0.5 & 0.3&   17.76&  1.07&   0.57\\
51&     $-$0.2 & 0.3&   $-$0.8 & 0.3&   18.16&  0.83&   0.48\\
52&     $-$0.3 & 0.4&   $ $~~0.0 & 0.5&   18.93&  0.89&   0.46\\

\hline
\end{tabular}

\small{$ ^{(a)}$Star IDs are the same as those in ALP}\\
\end{flushleft}  
\end{table*}

\begin{table*}
\begin{flushleft}
\begin{tabular}{cccccccc}  
\multicolumn{8}{c}{Table 6} \\ 
\multicolumn{8}{c} {Local Reference System for the Q0615$-$6615 field}\\
\hline \hline
Star$^{(a)}$&  $\mu_{\alpha}$cos$\delta$ & $\sigma$ & $\mu_{\delta}$ & $\sigma$ & V &B$-$V
 &V$-$R \\
ID &mas yr$^{-1}$&mas yr$^{-1}$ & mas yr$^{-1}$ & mas yr$^{-1}$ &mag & mag  
& mag \\  
\hline    

1&      $ $~~0.0 & 0.2&   $+$0.1 & 0.4&   18.95&  0.87&   0.53\\
2&      $ $~~0.0 & 0.2&   $+$0.7 & 0.3&   18.29&  0.83&   0.47\\
3&      $-$0.1 & 0.2&   $-$0.4 & 0.3&   17.46&  0.75&   0.43\\
4&      $ $~~0.0 & 0.3&   $-$1.0 & 0.4&   19.14&  0.61&   0.41\\
5&      $-$0.2 & 0.2&   $+$0.3 & 0.2&   18.23&  0.76&   0.45\\
6&      $+$0.4 & 0.3&   $+$0.9 & 0.3&   19.00&  0.98&   0.56\\
7&      $-$0.6 & 0.2&   $-$0.4 & 0.2&   19.07&  0.65&   0.42\\
8&      $+$0.7 & 0.2&   $+$0.1 & 0.2&   18.37&  0.89&   0.53\\
9&      $-$0.2 & 0.4&   $+$0.7 & 0.5&   18.98&  0.84&   0.49\\
10&     $-$0.2 & 0.3&   $-$0.9 & 0.4&   18.85&  0.85&   0.48\\
11&     $ $~~0.0 & 0.3&   $-$0.3 & 0.4&   18.97&  ....&   0.73\\
12&     $ $~~0.0 & 0.2&   $-$0.7 & 0.2&   18.25&  1.07&   0.53\\
13&     $-$0.1 & 0.2&   $ $~~0.0 & 0.3&   17.59&  1.04&   0.65\\
14&     $+$0.7 & 0.2&   $+$0.8 & 0.2&   18.33&  1.00&   0.57\\
15&     $+$0.4 & 0.4&   $-$0.6 & 0.5&   19.36&  0.90&   0.50\\
16&     $-$0.9 & 0.4&   $+$0.6 & 0.5&   19.29&  0.81&   0.47\\

\hline
\end{tabular}

\small{$ ^{(a)}$Star IDs are the same as those in ALP}\\
\end{flushleft}  
\end{table*}

\newpage

\begin{table*}
\begin{flushleft}
\begin{tabular}{lcccccccc}  
\multicolumn{7}{c}{Table 7} \\ 
\multicolumn{7}{c}{Mean barycentric positions of  quasars in the LMC}\\
\hline \hline
Epoch & \mbox{$\Delta\alpha$ cos$\delta$} & $\sigma$
&\mbox{$\Delta\delta$} 
& $\sigma$& N &CCD chip \\
 &arcsec&mas&arcsec&mas & & \\ 
\hline    
Q0459-6427:\\
1989.907&       8.443&  1.4&          $-$7.614&       1.1&    4&      RCA No.5\\
1990.872&       8.434&  0.3&          $-$7.615&       2.7&    3&      Tek No. 4\\
1990.878&       8.438&  5.8&          $-$7.624&       2.8&    2&      RCA No.5\\
1993.800&       8.423&  2.0&          $-$7.615&       2.6&    3&      Tek1024 No.1\\
1993.953&       8.432&  1.4&          $-$7.611&       0.7&    9&      Tek1024 No.2\\
1994.916&       8.429&  2.0&          $-$7.615&       1.1&    3&      Tek1024 No.2\\
1996.860&       8.421&  1.9&          $-$7.618&       2.0&    5&      Tek 2048 No.4\\
1998.881&       8.422&  0.8&          $-$7.615&       0.8&    6&      Tek1024 No.2\\
2000.010&       8.422&  0.4&          $-$7.618&       1.2&    9&      Tek1024 No.2\\
2001.961&       8.416&  2.1&          $-$7.612&       2.9&    3&      Tek1024 No.2\\
\hline 
Q0557-6713:\\
1989.024&       0.045&  1.6&          $-$2.768&       1.6&    5&      RCA No.5\\
1989.905&       0.039&  1.8&          $-$2.768&       1.9&    8&      RCA No.5\\
1990.872&       0.037&  1.6&          $-$2.772&       1.0&    4&      Tek No. 4\\
1990.878&       0.040&  2.7&          $-$2.766&       0.5&    3&      RCA No.5\\
1991.938&       0.046&  2.5&          $-$2.769&       0.7&    6&      Tek1024 No.1\\
1992.812&       0.042&  0.7&          $-$2.774&       0.6&    5&      Tek2048 No.1\\
1993.055&       0.040&  2.2&          $-$2.776&       1.3&    4&      Tek1024 No.1\\
1993.800&       0.039&  1.8&          $-$2.775&       0.7&    3&      Tek1024 No.1\\
1993.953&       0.039&  1.5&          $-$2.777&       1.1&    9&      Tek1024 No.2\\
1994.119&       0.033&  1.2&          $-$2.778&       0.9&    5&      Tek1024 No.2\\
1994.918&       0.036&  0.8&          $-$2.783&       0.7&    8&      Tek1024 No.2\\
1996.862&       0.033&  0.9&          $-$2.780&       0.3&    3&      Tek 2048 No.4\\
1998.883&       0.031&  0.3&          $-$2.786&       0.7&    6&      Tek1024 No.2\\
2001.961&       0.030&  0.3&          $-$2.785&       0.9&    3&      Tek1024 No.2\\
\hline
\end{tabular}
\end{flushleft}  
\end{table*}

\begin{table*}
\begin{flushleft}
\begin{tabular}{lcccccccc}  
\multicolumn{7}{c}{Table 7 (continued)} \\ 
\multicolumn{7}{c}{Mean barycentric positions of  quasars in the LMC}\\
\hline \hline
Epoch & \mbox{$\Delta\alpha$ cos$\delta$} & $\sigma$
&\mbox{$\Delta\delta$} & $\sigma$& N  &CCD chip \\
&arcsec&mas&arcsec&mas & & \\  
\hline    
Q0558-6707:\\
1992.813&       $-$12.148&      1.3&          $-$15.542&      1.3&    4&      Tek2048 No.1\\
1993.058&       $-$12.145&      0.8&          $-$15.534&      1.8&    4&      Tek1024 No.1\\
1993.953&       $-$12.148&      1.2&          $-$15.542&      3.4&    9&      Tek1024 No.2\\
1994.118&       $-$12.154&      1.6&          $-$15.541&      1.0&    6&      Tek1024 No.2\\
1994.918&       $-$12.149&      0.6&          $-$15.547&      0.9&    7&      Tek1024 No.2\\
1996.863&       $-$12.150&      1.6&          $-$15.547&      2.0&    6&      Tek 2048 No.4\\
1998.886&       $-$12.152&      0.7&          $-$15.540&      0.5&    3&      Tek1024 No.2\\
1999.942&       $-$12.159&      1.3&          $-$15.552&      1.3&    6&      Tek1024 No.2\\
2001.958&       $-$12.158&      1.1&          $-$15.541&      1.1&    6&      Tek1024 No.2\\
\hline 
Q0615-6615:\\
1989.908&       7.248&  1.3&          $-$8.255&       0.9&    3&      RCA No.5\\
1993.058&       7.242&     &          $-$8.266&          &    1&      Tek1024 No.1\\
1993.953&       7.244&  0.8&          $-$8.270&       2.1&    7&      Tek1024 No.2\\
1994.120&       7.238&  1.3&          $-$8.269&       3.9&    5&      Tek1024 No.2\\
1994.920&       7.236&  1.2&          $-$8.269&       1.4&    5&      Tek1024 No.2\\
1995.178&       7.234&  0.2&          $-$8.275&       3.3&    3&      Tek1024 No.2\\
1996.864&       7.234&  2.9&          $-$8.272&       1.5&    3&      Tek 2048 No.4\\
1997.194&       7.233&  1.7&          $-$8.274&       2.2&    5&      Tek1024 No.2\\
1998.886&       7.230&  1.5&          $-$8.276&       1.0&    3&      Tek1024 No.2\\
1999.942&       7.227&  1.2&          $-$8.278&       0.8&    3&      Tek1024 No.2\\
2001.960&       7.226&  1.3&          $-$8.277&       1.4&    12&     Tek1024 No.2\\

\hline
\end{tabular}
\end{flushleft}  
\end{table*}

\begin{table*}
\begin{flushleft}
\begin{tabular}{ccccccccc}
\multicolumn{6}{c}{Table 8} \\
\multicolumn{6}{c}{Proper Motion of the LMC (as measured)}\\
\hline \hline
Field &   $\mu_{\alpha}$cos($\delta$) & {$\mu_{\delta}$}&\# Frames&\ Epochs& Epoch Range    \\
ID & mas yr$^{-1}$  & mas yr$^{-1}$  &  \\
\hline

Q0459-6427&1.8 $\pm$  0.2&      0.1 $\pm$       0.2&    47&     9&      1989.91$-$2001.96\\
Q0557-6713&1.1 $\pm$  0.2&      1.9 $\pm$       0.1&    72&     13&     1989.02$-$2001.96\\
Q0558-6707&1.2 $\pm$  0.2&      0.6 $\pm$       0.3&    51&     9&      1992.81$-$2001.96\\
Q0615-6615&1.9 $\pm$  0.2&      1.4 $\pm$       0.2&    50&     11&     1989.90$-$2001.96\\

\hline
\end{tabular}
\end{flushleft}
\end{table*}

\begin{table*}     
\begin{flushleft}     
\begin{tabular}{lcccc}     
\multicolumn{4}{c}{Table 9} \\     
\multicolumn{4}{c}{High precision determinations of the proper motion of the LMC}\\  

\noalign{\smallskip}     
\hline \hline     
\noalign{\smallskip}     
~~~~{Source}& $\mu_{\alpha}$cos($\delta$) & {$\mu_{\delta}$}& {Proper Motion System}    \\     
 & mas yr$^{-1}$  & mas yr$^{-1}$  &  \\     
\noalign{\smallskip}     
\hline     
\noalign{\smallskip}

Kroupa, R\"{o}ser \&  Bastian 1994 (Field) &$+$1.3 $\pm$ 0.6 &$+$1.1 $\pm$ 0.7
                &PPM\\
Jones et al. 1994       &$+$1.37 $\pm$ 0.28 &$-$0.18 $\pm$ 0.27 &Galaxies\\
Kroupa \& Bastian 1997 (Field)  &$+$1.94 $\pm$ 0.29 &$-$0.14 $\pm$ 0.36 &Hipparcos\\
ALP &$+$1.7 $\pm$ 0.2 &$+$2.9 $\pm$ 0.2 &Quasars\\
PAM &$+$2.0 $\pm$ 0.2 &$+$0.4 $\pm$ 0.2 &Quasars\\
Drake et al. 2001 &$+$1.4 $\pm$ 0.4 &$+$0.38 $\pm$ 0.25 &Quasars\\
Kallivayalil et al. 2005 &$+$2.03 $\pm$ 0.08 &$+$0.44 $\pm$ 0.05 &Quasars\\
This work (Field)\tablenotemark{a} &$+$1.5 $\pm$ 0.1 &$+$1.4 $\pm$ 0.1 &Quasars\\
This work\tablenotemark{a} &$+$1.8 $\pm$ 0.1 &$+$0.9 $\pm$ 0.1 &Quasars\\

\noalign{\bigskip}           
\hline      
\end{tabular}     

\small{$^{(a)}$ Weighted mean of our four QSO fields} \\
\end{flushleft}     
\end{table*}

\begin{table*}     
\small
\begin{flushleft}     
\begin{tabular}{lrrrr}     
\multicolumn{5}{c}{Table 10} \\     
\multicolumn{5}{c}{Proper Motion and Space Velocity Results for the LMC}\\     
\noalign{\smallskip}     
\hline \hline     
\noalign{\smallskip}     
~~~~{Parameter} & {Q0459-6427} & {Q0557-6713} & {Q0558-6707} & {Q0615-6615} \\     
\noalign{\smallskip}     
\hline          
\noalign{\smallskip}     
$\Delta \mu_{\alpha}\cos{\delta}$, rotation correction (mas yr$^{-1}$)     
&$+$0.17 &$+$0.11      &$+$0.11        &$+$0.12\\
\noalign{\smallskip}     
$\Delta \mu_{\delta}$, rotation correction (mas yr$^{-1}$)     
&$+$0.09& $-$0.18       & $-$0.18& $-$0.18      \\         
\noalign{\smallskip}     
$\mu^{LMC}_{\alpha}\cos{\delta}$, LMC centered (mas yr$^{-1}$)  
& 1.9 $\pm$ 0.2 & 1.5 $\pm$ 0.2 & 1.4 $\pm$ 0.2 & 2.2 $\pm$ 0.2  \\     
\noalign{\smallskip}     
$\mu^{LMC}_{\delta}$, LMC centered (mas yr$^{-1}$)  
& 0.5 $\pm$ 0.2 & 1.5 $\pm$ 0.1 & 0.2 $\pm$ 0.2 & 0.7 $\pm$ 0.2  \\     

\noalign{\smallskip}     
$\mu^{GRF}_{\alpha}\cos{\delta}$ (mas yr$^{-1}$)  
& 1.4 $\pm$ 0.1 & 1.0 $\pm$ 0.1 & 0.9 $\pm$ 0.1 & 1.7 $\pm$ 0.1  \\     
\noalign{\smallskip}     
$\mu^{GRF}_{\delta}$ (mas yr$^{-1}$)  
& 0.4 $\pm$ 0.2 & 1.3 $\pm$ 0.1 & 0.1 $\pm$ 0.3 & 0.6 $\pm$ 0.2  \\     
    
\noalign{\smallskip}     
$\mu^{GRF}_{l}\cos{b}$ (mas yr$^{-1}$)
& $-$0.6 $\pm$ 0.2 & $-$1.5 $\pm$ 0.1 & $-$0.3 $\pm$ 0.3 & $-$0.9 $\pm$ 0.2  \\     

\noalign{\smallskip}     
$\mu^{GRF}_{b}$ (mas yr$^{-1}$)
& 1.4 $\pm$ 0.1 & 0.7 $\pm$ 0.1 & 0.8 $\pm$ 0.1 & 1.5 $\pm$ 0.1  \\     

\noalign{\smallskip}     
$\Pi$, velocity component (km s$^{-1}$)     
& $ $252 $\pm$ 25 & $ $215 $\pm$ 23 & $ $171 $\pm$ 28 & $ $292 $\pm$ 23  \\     

\noalign{\smallskip}     
$\Theta$, velocity component (km s$^{-1}$)     
& $ $93 $\pm$ 41 & $ $319 $\pm$ 31 & $ $27 $\pm$ 63 & $ $160 $\pm$ 45  \\     

\noalign{\smallskip}     
$Z$, velocity component (km s$^{-1}$) 
& 234 $\pm$ 25 & 109 $\pm$ 24 & 135 $\pm$ 26 & 274 $\pm$ 22  \\     

\noalign{\smallskip}     
V$_{\rm {gc, r}}$, radial velocity (km s$^{-1}$) 
& 80 $\pm$ 23  &118 $\pm$ 22   &68 $\pm$  24  &92 $\pm$  20  \\     

\noalign{\smallskip}     
V$_{\rm {gc, t}}$, transverse velocity (km s$^{-1}$) 
& 347 $\pm$ 27  &382 $\pm$ 30   &209 $\pm$  27  &421 $\pm$  27  \\     
\noalign{\smallskip}     
\hline     
\noalign{\bigskip}     
\end{tabular}     
     
\end{flushleft}     
\end{table*}

\begin{table*}     
\begin{flushleft}     
\begin{tabular}{lcccc}     
\multicolumn{5}{c}{Table 11} \\     
\multicolumn{5}{c}{Mass of the Galaxy for three LMC rotational velocities}\\    

\noalign{\smallskip}     
\hline \hline     
\noalign{\smallskip}     
{Parameter} & {Q0459-6427}& {Q0557-6713}& {Q0558-6707} & {Q0615-6615}\\     
\noalign{\smallskip}     
\hline      
\noalign{\smallskip}
v$_{\Phi}$ = 50 km s$^{-1}$  &   \\
V$_{\rm {gc, r}}$, radial velocity (km s$^{-1}$)
&80 $\pm$ 23 &118 $\pm$ 22 &68 $\pm$ 24 &92 $\pm$ 20 \\
\noalign{\smallskip}
V$_{\rm {gc, t}}$, transverse velocity (km s$^{-1}$)
&347 $\pm$ 27 &382 $\pm$ 30 &209 $\pm$ 27 &421 $\pm$ 27 \\
M$_{\rm G}$, mass of the Galaxy in $10^{11}\times \cal M_{\sun}$
& (8.2 $\pm$ 1.3) & (9.9 $\pm$ 1.6) & (3.0 $\pm$ 0.8) & (12 $\pm$ 2) \\
\noalign{\smallskip}
\hline     
v$_{\Phi}$ = 0 km s$^{-1}$  &   \\          
V$_{\rm {gc, r}}$, radial velocity (km s$^{-1}$) 
&75 $\pm$ 23 &126 $\pm$ 22 &75 $\pm$ 24 &99 $\pm$ 20 \\     
\noalign{\smallskip}     
V$_{\rm {gc, t}}$, transverse velocity (km s$^{-1}$) 
&305 $\pm$ 26 &408 $\pm$ 30 &198 $\pm$ 33 &420 $\pm$ 30 \\     
M$_{\rm G}$, mass of the Galaxy in $10^{11}\times \cal M_{\sun}$ 
& (6.3 $\pm$ 1.1) & (11 $\pm$ 2) & (2.7 $\pm$ 0.9) & (12 $\pm$ 2) \\        
\noalign{\smallskip}
\hline      
v$_{\Phi}$ = 90 km s$^{-1}$ & \\         
V$_{\rm {gc, r}}$, radial velocity (km s$^{-1}$) 
&83 $\pm$ 23 &112 $\pm$ 22 &62 $\pm$ 24 &86 $\pm$ 20 \\     
\noalign{\smallskip}     
V$_{\rm {gc, t}}$, transverse velocity (km s$^{-1}$) 
&381 $\pm$ 27 &364 $\pm$ 29 &225 $\pm$ 26 &425 $\pm$ 25 \\     
\noalign{\smallskip}     
M$_{\rm G}$, mass of the Galaxy in $10^{11}\times \cal M_{\sun}$ 
& (9.9 $\pm$ 1.4) & (9.0 $\pm$ 1.5) & (3.4 $\pm$ 0.8) & (12 $\pm$ 2) \\
\noalign{\smallskip}     
\hline 
\end{tabular}     
\end{flushleft}     
\end{table*}

\end{document}